\def\beq{\begin{equation}}
\def\eeq{\end{equation}}
\def\beqa{\begin{eqnarray}}
\def\eeqa{\end{eqnarray}}
\def\ua{\uparrow}
\def\da{\downarrow}
\newcommand{\cuo}{CuO$_2$ }
\newcommand{\hg}{HgBa$_2$Ca$_4$Cu$_5$O$_y$ }
\newcommand{\iu}{{\rm i}}
\newcommand{\lco}{La$_2$CuO$_4$ }
\newcommand{\lsco}{La$_{1.85}$Sr$_{0.15}$CuO$_4$ }
\newcommand{\para}{\parallel}
\newcommand{\rmafi} {{\rm AFI}}
\newcommand{\rmd}{{\rm d}}
\newcommand{\rme}{{\rm e}}

\newcommand{\rmsc} {{\rm SC}}
\newcommand{\tl}{TlBa$_2$Ca$_4$Cu$_5$O$_y$ }
\newcommand{\tc} {$T_{\rm c}$}
\newcommand{\tzero} {$T_{\rm 0}$}
\newcommand{\tr} {{\rm Tr}}
\newcommand{\etal} {{\it et al.}}
\newcommand{\pconf}{``$\pi$-config''}
\newcommand{\zconf}{``0-config''}
\def\jnl#1#2#3#4{#1 {\bf #2}, #3 (#4).}
\def\EPJB{Eur.\ Phys.\ J. B}

\def\PRB{Phys.\ Rev.\ B}

\documentclass[prl,twocolumn,amsmath,amssymb,showpacs]{revtex4}
\usepackage{graphicx}
\begin{document}
\title{Effects of antiferromagnetic planes on the superconducting properties of multilayered high-$T_{\rm c}$ cuprates}
\author{M. Mori and S. Maekawa} 
\affiliation{
	Institute for Materials Research, Tohoku University, Sendai 980-8577, Japan}
\date{\today}
\begin{abstract}
We propose a mechanism for high critical temperature ($T_c$) in the coexistent phase of superconducting- (SC) and antiferromagnetic (AF) CuO$_2$ planes in multilayered cuprates. 
The Josephson coupling between the SC planes separated by an AF insulator (Mott insulator) is calculated perturbatively up to the fourth order in terms of the hopping integral between adjacent CuO$_2$ planes. 
It is shown that the AF exchange splitting in the AF plane suppresses the so-called {\it $\pi$-Josephson coupling}, and the long-ranged {\it 0-Josephson coupling} leads to coexistence with a rather high value of $T_c$.

\end{abstract}
\maketitle
There is considerable interest in the superconducting critical temperature, $T_{\rm c}$, in cuprates. 
In multilayered cuprates having several \cuo planes in a conducting block, \tc~increases with the number of \cuo planes, $n$, and has a maximum at $n$=3 \cite{karppinen}. 
Several studies have proposed that the suppression of \tc~for $n>3$ is caused by a charge imbalance among individual \cuo planes \cite{kotegawa,jansen,mori,chak_nat}; 
the outer-pyramidal-coordinated-planes (OP's) tend to get optimal- or overdoped, while the inner-square-coordinated-planes (IP's) tend to get underdoped \cite{karppinen,trokiner,kotegawa}. 
Chakravarty $et~al.$ have claimed that a Josephson coupling enhances the \tc~up to $n=3$, whereas  a sizeable charge imbalance combined with competing order parameters reduces \tc~beyond $n=3$ \cite{chak_nat}.

Recently, a coexistence of superconducting- (SC) and antiferromagnetic (AF) states has been observed in five-layered cuprates, \hg and \tl\cite{kotegawa2}, and in a heterostructure composed of alternating stack of \lsco and \lco\cite{bosovic}.  
In the five-layered cuprates, since the charge imbalance is enhanced by increasing  $n$\cite{kotegawa}, the underdoped IPs and the optimally doped OPs show AF- and SC states, respectively\cite{kotegawa2}. 
It is noted that the five-layered cuprates retain rather high values of \tc=100$\sim$108 K\cite{kotegawa}, despite the fact that the SC planes are separated by AF planes in the direction perpendicular to the planes.  
In general, a Josephson coupling between SC planes is necessary both to stabilize the bulk SC state and to enhance \tc~in layered superconductors\cite{chak_nat,anderson,tesanovic,leggett,chak_EPJB}. 
Therefore, in the above coexistent phases,  in the five-layered cuprates, the Josephson coupling is required via AF planes not only for the stability of the superconductivity but also for such high values of \tc.

In this Letter, we study the coexistence of SC- and AF \cuo planes in multilayered cuprates. 
The AF plane is assumed to be an insulator at half-filling with no double occupancy.
The Josephson coupling between SC planes separated by AF one is perturbatively calculated in terms of the hopping integral between adjacent \cuo planes.  
The perturbative processes comprises two parts: The first provides a positive value of Josephson coupling called {\it 0-Josephson coupling}, while the second makes a negative contribution called {\it $\pi$-Josephson coupling}. 
Note that the sign of Josephson coupling reflects a quantum effect originating from the fermion anticommutation rules\cite{glatzman,spivak,choi}. 
We find that the AF exchange interaction suppresses the latter process, and allows the Cooper pair to tunnel through 
the AF insulating (AFI) plane.
The fluctuations of the SC phase are suppressed by this long ranged Josephson coupling, and it is this which enables the coexistence and a rather high value of \tc. 
The $n$-dependence of \tc~and enhancement of the Josephson coupling are discussed. 

\begin{figure}[b]
	\includegraphics[height=3cm]{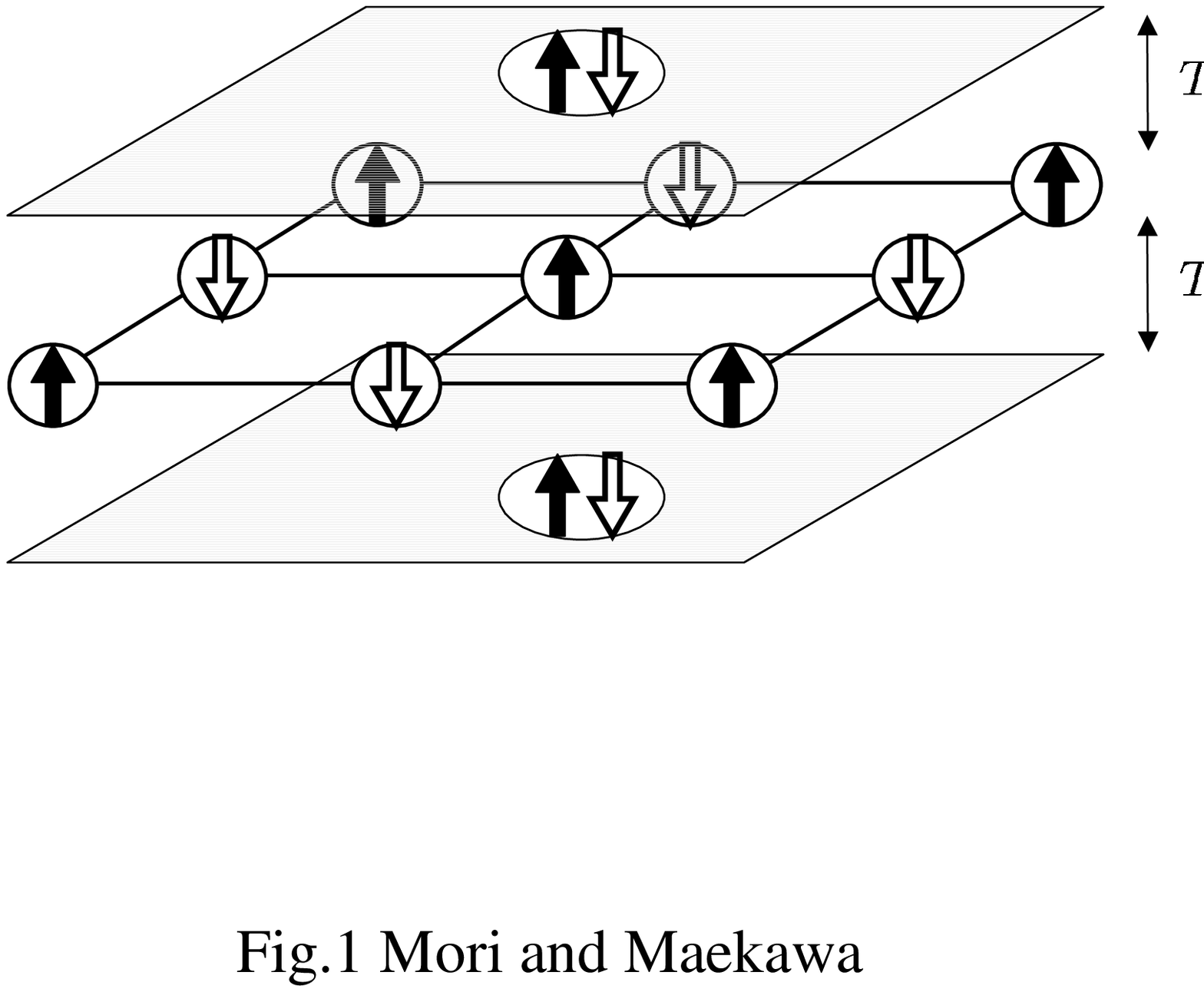}
\caption{Schematic figure of a conducting block in the three-layered system, SC/AFI/SC. 
The thick and open arrows indicate spins of electrons. 
The up- and down spins enclosed by the oval indicate condensed Cooper pairs. 
The matrix element of interlayer hopping is denoted by $T_k$.}
\label{fig1}
\end{figure}
The minimal model is a three-layered system composed of two SC planes with $d$-wave symmetry and an AFI plane at half-filling. 
The SC planes are separated by the AFI plane as shown in Fig.~\ref{fig1}. 
In the five-layered cuprates, the two SC planes are separated by three AFI planes, and the same mechanism arises in higher order. 
Coexistence in five-layered cuprates is explained by the Josephson coupling through the AFI planes.

In each SC plane, the BCS mean-field Hamiltonian is adopted, and the wavefunctions in the two SC planes are given by
\beqa
|\rmsc_1\rangle
	&\equiv&
		\prod_{k} (u_{k}+v_{k}\rme^{\iu\phi_1}a_{k\ua}^{\dag}a_{-k\da}^{\dag})|0\rangle,\\
|\rmsc_2\rangle
	&\equiv&
		\prod_{k} (u_{k}+v_{k}\rme^{\iu\phi_2}c_{k\ua}^{\dag}c_{-k\da}^{\dag})|0\rangle,
\eeqa
where $v_k/u_k=\Delta_k/E_k$ and $E_k = \sqrt{\xi_k^2+\Delta_k^2}$. 
The SC order parameter is denoted by $\Delta_k = (\Delta_0/2)[\cos(k_x)-\cos(k_y)]$ and $\xi_k$ is the quasiparticle energy in the normal state. 
The true vacuum is indicated by $|0\rangle$. 
The operators, $a_{k\sigma}^{\dag}$ and $c_{k\sigma}^{\dag}$, create electrons with momentum, $k$, and spin, $\sigma$, in the SC$_1$ and the SC$_2$, respectively.

In the AFI plane, the interaction between localized spins is given by $J\sum_{\langle i,j\rangle} \vec{\rm S}_i\cdot\vec{\rm S}_j$, where $\vec{\rm S}_i$ is the spin operator at $i$-th site, and the summation runs over the nearest neighbor sites. 
The N\'eel state is assumed for the ground state, and its wavefunction is given by
\beqa
|\rmafi\rangle
	&\equiv&
	  \prod_{i\in A,j\in B}b_{i\ua}^{\dag}b_{j\da}^{\dag}|0\rangle,
\label{afi}
\eeqa
where $b_{i\sigma}^{\dag}$ is the electron creation operator at $i$-th site with  spin $\sigma$.  Up- and down spins are sited on sublattices $A$ and $B$, respectively. 
The phase convention is defined by putting the operators in order of its site index.
No double occupancy is imposed on $|\rmafi\rangle$. 
The charge imbalance between the SC and the AFI planes is induced by a site potential, $W$, whose value is of the order of $J$\cite{mori}. 
Due to this potential, intermediate states with a double occupancy are higher in energy than those with a single hole in the multilayered cuprates\cite{kotegawa2}. 
\begin{figure*}[t]
	\includegraphics[height=5cm]{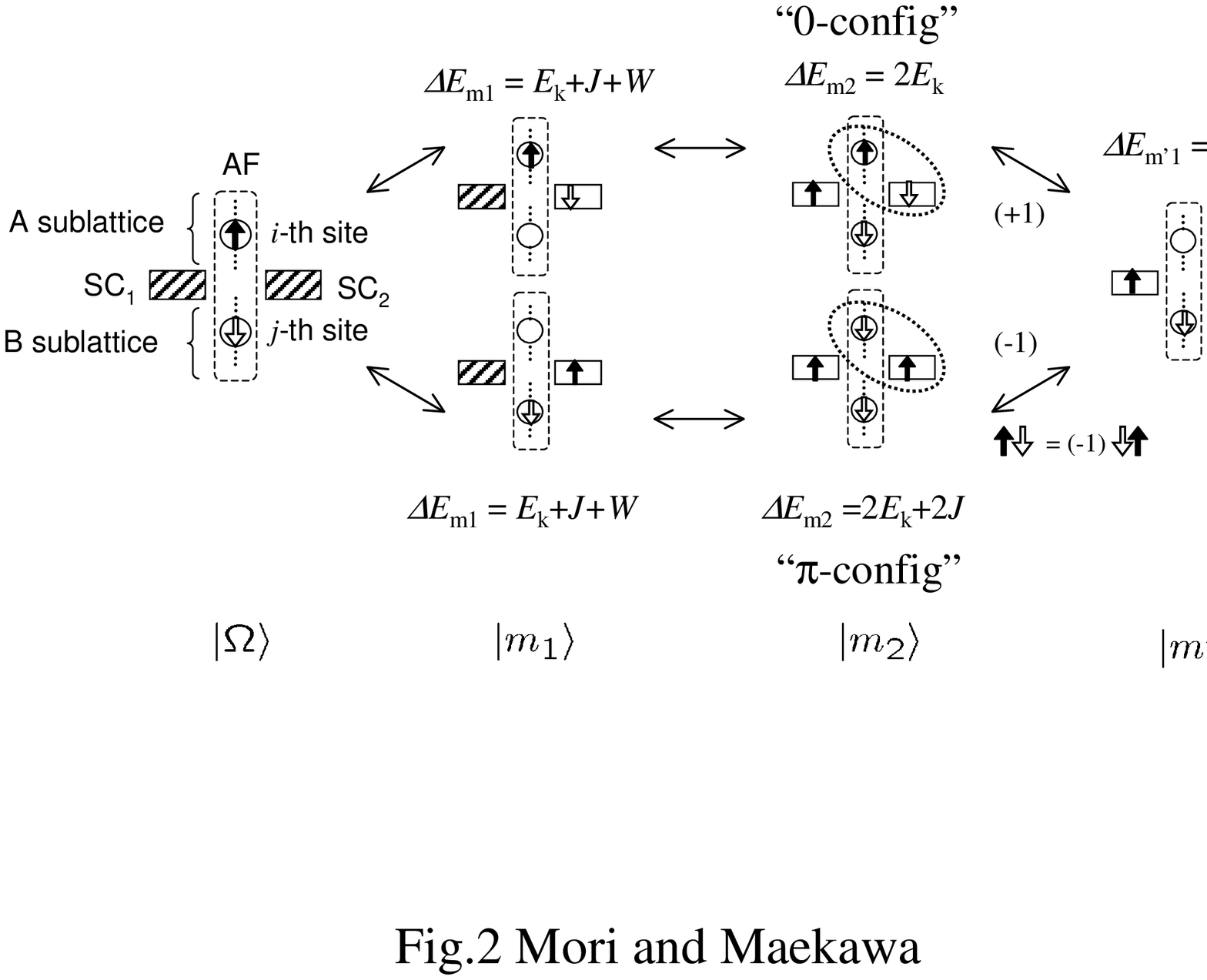}
\caption{Two examples of tunneling processes contributing to the Josephson coupling energy,  $E_J= \sum
		\langle\Omega|H_T|m_1\rangle
		\langle m_1|H_T|m_2\rangle
		\langle m_2|H_T|m_1'\rangle
		\langle m_1'|H_T|\Omega\rangle/({\it \Delta}E_{m_1}{\it \Delta}E_{m_2}{\it \Delta} E_{m_1'})$. 
The upper flow provides the {\it 0-Josephson coupling}, while the lower flow leads to the {\it $\pi$-Josephson coupling}. The shaded rectangles imply the SC ground state. The arrows in the SC and in the AF indicate the quasiparticle excitations and the localized spins, respectively. 
The open circle denotes a vacant site in the AFI plane. 
The \pconf~ has the parallel spin configuration in the SCs, while the \zconf~ has the anti-parallel configuration. 
The anticommutation of fermions occurs only in the lower flow. }
\label{fig2}
\end{figure*}

The AF- and the SC planes are connected by the tunneling Hamiltonian as
\beqa
H_T &=&
	H_1 + H_2,\label{tunnel}\\
H_1
	&=&
		\sum_{i,k,\sigma} 
			\left(\phi_{i,k}   a^{\dag}_{k\sigma}b_{i\sigma}
				   +\phi^*_{i,k}b^{\dag}_{i\sigma} a_{k\sigma}
			\right),\\
H_2
	&=&	
		\sum_{i,k,\sigma} 
			\left(\phi_{i,k}   b^{\dag}_{i\sigma}c_{k\sigma}
				   +\phi^*_{i,k} c^{\dag}_{k\sigma}b_{i\sigma}
			\right),\\
\phi_{i,k}
	&=&
		\frac{1}{N^{1/2}}\rme^{-\iu k r_i}T_k,
\eeqa
where an electron coherently hops between adjacent planes with matrix element, $T_k=t_{\perp}/4\left(\cos(k_x)-\cos(k_y)\right)^2$\cite{chak_sci,okanderson,liechtenstein}, and $b_{i\sigma}=1/N^{1/2}\sum_k \rme^{\iu k r_i} b_{k\sigma}$.

The Josephson coupling energy, $-E_J\cos\theta$, which is a function of phase difference between SC$_1$ and SC$_2$, $\theta\equiv\phi_1-\phi_2$, is obtained by the fourth order perturbation theory in terms of Eq.~(\ref{tunnel}). 
The wavefunction of the ground state is given by
$
|\Omega\rangle
	=
		|\rmsc_1\rangle\otimes |\rmafi\rangle\otimes |\rmsc_2\rangle,\label{sas}
$
where the order of $|\rmsc_1\rangle$, $|\rmafi\rangle$, and $|\rmsc_2\rangle$ must be maintained to define a phase convention. 

The first intermediate states, $|m_1\rangle$ and $|m'_1\rangle$, are obtained by transferring an electron from the AFI plane to the SC one as shown in Fig.~\ref{fig2}, since the double occupancy is forbidden in $|\rmafi\rangle$. 
The energy of $|m_1\rangle$ and $|m'_1\rangle$ is given by ${\it \Delta} E_{m_1} ={\it \Delta} E_{m_1'} = E_k+J+W$, where $W$ is the site potential in the IP\cite{mori}. Spin fluctuations and hole motions are neglected. 

The second intermediate states, $|m_2\rangle$, that can provide the Josephson coupling energy, is classified into two types of spin configurations, i.e., \zconf~ and \pconf. 
Typical processes are shown in Fig.~\ref{fig2}. 
Each SC plane has one quasiparticle excitation, and the AFI plane has no hole.  
The \zconf~ has an antiparallel spin configuration in the SC planes, while the \pconf~ has a parallel one.  
The energy of $|m_2\rangle$ with \zconf~ is given by ${\it \Delta} E_{m_2} =2E_k$, where the spin configuration in the AFI plane is the same as that in the ground state. 
On the other hand, the energy of $|m_2\rangle$ with \pconf~ is given by ${\it \Delta} E_{m_2} =2E_k+2J$, since one site is filled with an opposite spin. 

Finally, we find that the magnitude of Josephson coupling energy is given by
\beqa
E_J
	&=&
		E_J^0+E_J^{\pi},\label{ej_afi}\label{ejsum}\\
	&\sim &
		\left(\frac{1}{\Delta_0}-\frac{1}{(\Delta_0+J)}\right)\frac{t_{\perp}^4}{(\Delta_0+J+W)^2},\label{ej}\\
E_J^0
	&=&
	4\sum_k  \frac{T_k^4}{2E_k(E_k+J+W)^2}\left(\frac{\Delta_k}{2 E_k}\right)^2\label{ej0}\\
E_J^{\pi}
	&=&
	-4\sum_k  \frac{T_k^4}{(2E_k+2J)(E_k+J+W)^2}\left(\frac{\Delta_k}{2 E_k}\right)^2.\label{ejp}
\eeqa
Equations (\ref{ej0}) and (\ref{ejp}) are caused by \zconf~and \pconf, respectively. 
We look more carefully into the signs of $E_J^0$ and $E_J^{\pi}$. 
In the transitions from $|\Omega\rangle$ to $|m_2\rangle$ by way of $|m_1\rangle$, both \zconf~and \pconf~have the same sign. 
On the other hand, only in \pconf, the anticommutation of fermions occurs between $|m_2\rangle$ and $|m'_1\rangle$, and thus the additional minus sign is added to its transition amplitude. 
As a consequence, {\it the \zconf~provides the {\it 0-Josephson coupling}, while the \pconf~does the {\it $\pi$-Josephson coupling}.}  
{\it The signs of $E_J^0$ and $E_J^{\pi}$ are attributed to the quantum effect originating from the anticommutation of fermions}\cite{glatzman,spivak,choi}. 

\begin{figure}[b]
	\includegraphics[height=4.50cm]{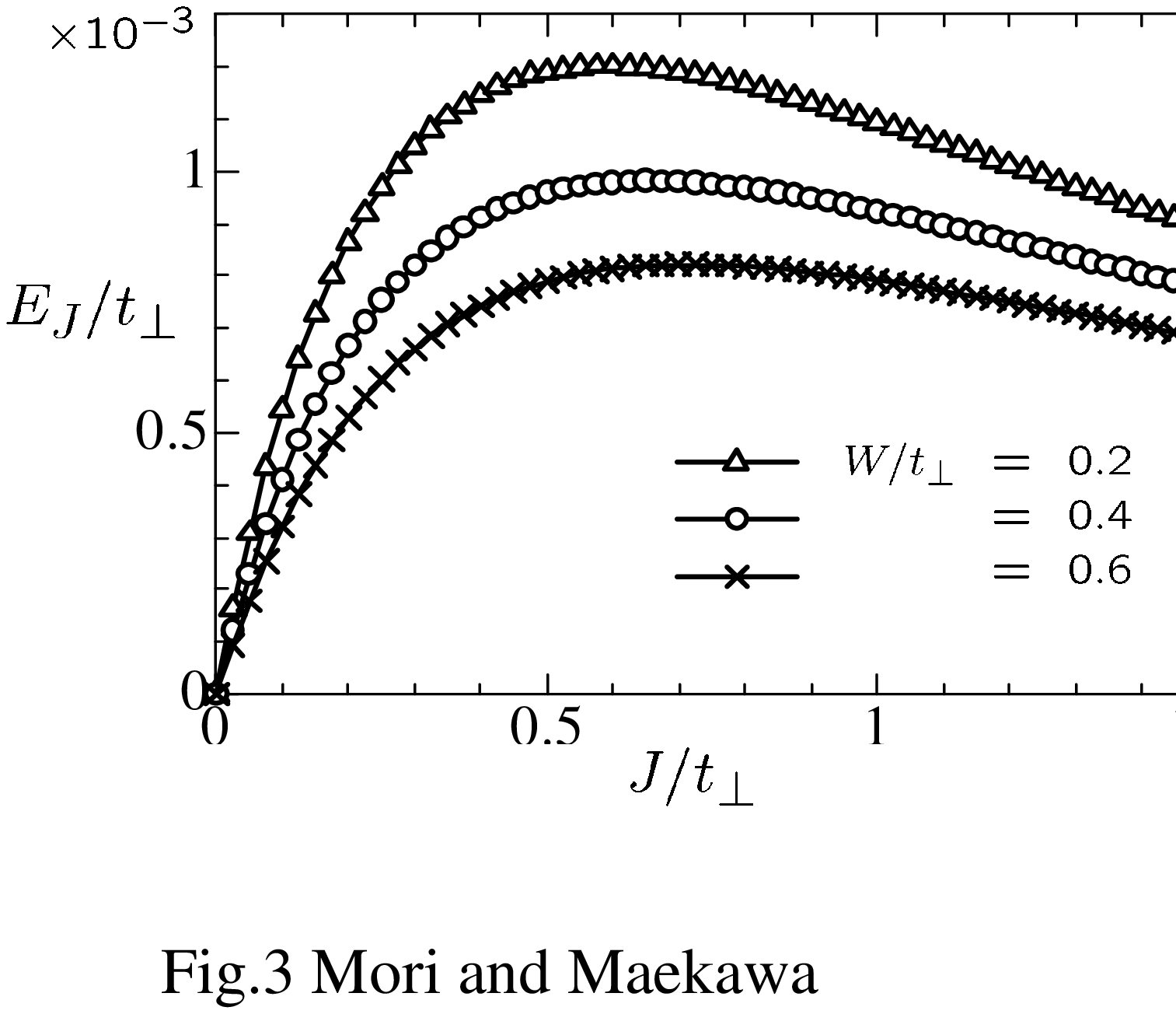}
\caption{$J$-dependence of $E_J$. We adopted $\xi_k=2t(\cos(k_x)+\cos(k_y))-\mu$ for the quasiparticle energy. Parameters are scaled by $t_{\perp}$ as $t/t_{\perp}=5$, $\Delta_0/t_{\perp}=1$, $\mu/t_{\perp}=-1$.}
\label{fig3}
\end{figure}
Note that, when the AF interaction in the AFI plane is much smaller than the SC gap, i.e., $J\ll \Delta_0$, the Cooper pair cannot go through the AFI plane, since \zconf~and \pconf~processes in Eq.~(\ref{ej}) cancel out as
\beqa
E_J
	&\sim&
		\left(\frac{1}{\Delta_0}-\frac{1}{(\Delta_0+0)}\right)\frac{t_{\perp}^4}{(\Delta_0+0+W)^2}=0.
\eeqa
To show the $J$-dependence of Eq.~(\ref{ejsum}), $E_J$ is numerically calculated and plotted in Fig.~\ref{fig3} as a function of $J/t_{\perp}$ for $W/t_{\perp}=0.2,~0.4$, and 0.8. 
We adopt $\xi_k=2t(\cos(k_x)+\cos(k_y))-\mu$ for the quasiparticle energy, and $t/t_{\perp}=5$, $\Delta_0/t_{\perp}=1$, and $\mu/t_{\perp}=-1$. 
One can find that {\it the AF interaction generates the Josephson coupling through the AFI plane.} 

We have shown that the long-ranged Josephson coupling through the AFI, $E_J$, can survive due to the magnetic exchange interaction. Although a magnitude of $E_J$ is small, it is important to provide the phase coherence, which plays an important role to determine $T_c$ in the cuprate superconductors\cite{emery,zaleski}. Below, we study the SC phase coherence in the multilayered systems based on a model proposed by Zaleski and Kope\'c\cite{zaleski}. In the present study, we take account of the long-ranged Josephson coupling denoted by $K$, that is crucial to obtain a rather high value of $T_c$ in the coexistent phase in the five-layered cuprates.

\begin{figure}[b]
	\includegraphics[height=4.30cm]{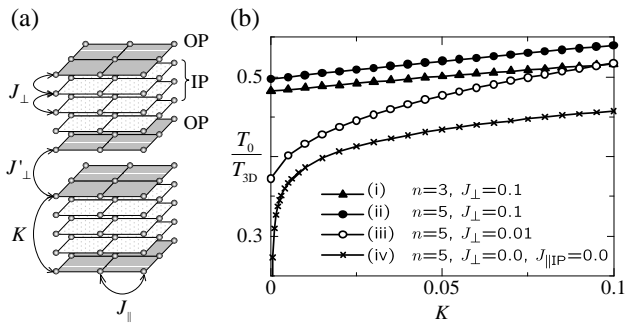}
\caption{(a) Schematic figure of five-layered XY-model.  
(b) $K$-dependence of $T_0$ normalized by that of the isotropic case, $T_{\rm 3D}$, given by $J_{\para{\rm OP}}$=$J_{\para{\rm IP}}$=$J_{\perp}$=$J'_{\perp}$=1 and $K$=0. Each line is give by as follows: 
(i) $n$=3, $J_{\para{\rm OP}}$=$J_{\para{\rm IP}}$=1, $J_{\perp}$=0.1, and $J'_{\perp}$=0.01 for triangle; 
(ii) $n$=5, $J_{\para{\rm OP}}$=$J_{\para{\rm IP}}$=1, $J_{\perp}$=0.1, and $J'_{\perp}$=0.01 for solid circle; 
(iii) $n$=5, $J_{\para{\rm OP}}$=$J_{\para{\rm IP}}$=1, $J_{\perp}$=0.01, and $J'_{\perp}$=0.01 for open circle; 
(iv) for cross, $n$=5 is reduced to $n$=2 by assuming $J_{\para{\rm OP}}$=1, $J_{\para{\rm IP}}$=0, $J_{\perp}$=0, and $J'_{\perp}$=0.01.}
\label{fig4}
\end{figure}

The free energy given by a spatial variation of the SC order parameter, $\Psi(r)$, is proportional to $\int\rmd r |\nabla \Psi(r)|^2\sim|\Psi_0|^2\int\rmd r (\nabla\phi(r))^2\sim|\Psi_0|^2\int\rmd r \cos(\phi_i-\phi_j)$.
We assume that the amplitude, $\Psi_0$, is constant and the spatial variation of phase, $\phi(r)$, is slow, i.e., $\phi_i-\phi_j\sim\nabla\phi(r)$. 
Therefore, the phase degree of freedom in the multilayered cuprates is given by the XY-model as
\beqa
H		&=&
	H_0 + H_1,\label{xymodel}\\
H_0	&=&
	- \sum_{\langle i,j\rangle,l} 
		J^{(\alpha)}_{\para} \; \vec{R}^{(\alpha)}_{i,l}\cdot\vec{R}^{(\alpha)}_{j,l}\nonumber\\
	&-& \!\!\!\!\!\! \sum_{i,l,\langle \alpha\beta\rangle} 
		J_{\perp} \; \vec{R}^{(\alpha)}_{i,l}\cdot\vec{R}^{(\beta)}_{i,l}
	-\!\!\!\!\!\! \sum_{i,\langle l,m \rangle,\langle \alpha\beta\rangle} \!\!\!\!\!\!
		J'_{\perp} \; \vec{R}^{(\alpha)}_{i,l}\cdot\vec{R}^{(\beta)}_{i,m},\\
H_1
	&=&
	- \sum_{i,l,\langle\langle \alpha\beta\rangle\rangle} 
		K^{(\alpha\beta)} \; \vec{R}^{(\alpha)}_{i,l}\cdot\vec{R}^{(\beta)}_{i,l},
\eeqa
where $\vec{R}^{(\alpha)}_{i,l}=(R_{i,l}^{(\alpha),x},R_{i,l}^{(\alpha),y})$ is the $XY$-spin operator at $i$-th site on the $\alpha$-th plane in the $l$-th conducting block. 
The single square brackets indicate sums between nearest neighboring sites, planes and blocks. The double square bracket indicates a sum between the OPs in one block. 
Schematic figure of planes are shown in Fig.~\ref{fig4} (a). 
The SC planes have a finite value of $J^{(\alpha)}_{\para}$, while $J^{(\alpha)}_{\para}=0$ in the AF planes. 
The long-ranged Josephson coupling via the AFI plane is denoted by $K$. 
The $J'_{\perp}$ connects the SC OP in one block to that in another block. 
If the IP is also the SC state, $J_{\perp}$ should be included between the OP and the IP within the block. Such a case is used to discuss the $n$-dependence of \tc.

The free energy par site for Eq.~(\ref{xymodel}) is given by
$f(\zeta)=-\zeta/\beta+2/(\beta N)\sum_{k,p}^{N/2} \tr\ln\left[\zeta-\beta\hat{M}_n \right]$, where we adopted an approximation that the average length of spins is restricted to 1\cite{zaleski}. 
The matrix, $\hat{M}_n$, is the Fourier transform of the Hamiltonian, Eq.~(\ref{xymodel}), in a $n$-layered system.
The Lagrange multiplier, $\zeta$, is determined by a saddle point equation as 
$\zeta_0-\beta_cE^{(1)}_{0}=0$, 
where 
$E^{(1)}_{0}>E^{(2)}_{0}>...>E^{(n)}_{0}$ 
are the eigenvalues of $\hat{M_n}$ at $k=0$. 

The phase coherence develops below the critical temperature, $T_0$, which is determined by 
\beq
T_0	=
	E^{(1)}_{0}
	\Bigl(\frac{1}{N}\sum_{k}^{N}\frac{1}{n} \sum_{\alpha=1}^{n}\frac{1}{1-E^{(\alpha)}_{k}/E^{(1)}_{0}}\Bigr)^{-1}. 
	\label{tc}
\eeq
When all inter-layer couplings are zero, i.e., $J_{\perp}=J_{\perp}'=K=0$, the $k$-summation in Eq.~(\ref{tc}) diverges, and then \tzero=0. 

The $K$-dependence of \tzero~is shown in Fig.~\ref{fig4} (b). 
\tzero~is normalized by $T_{\rm 3D}$, which denotes the critical temperature in the isotropic case on the three-dimensional cubic lattice, i.e., $J_{\para{\rm OP}}$=$J_{\para{\rm IP}}$=$J_{\perp}$=$J'_{\perp}$=1 and $K$=0. The ratio of \tzero~to $T_{\rm 3D}$ measures an effect of the interlayer couplings. 
The three- and five-layered cases with $J_{\para{\rm OP}}$=$J_{\para{\rm IP}}$=1, $J_{\perp}$=0.1, and $J'_{\perp}$=0.01, are plotted by solid circles and triangles, respectively. 
We find that \tzero~increases with $n$\cite{zaleski}. 
For the small value of $J_{\perp}$=0.01 in the five-layered case with $J_{\para{\rm OP}}$=$J_{\para{\rm IP}}$=1, and $J'_{\perp}$=0.01, \tzero~is suppressed as shown by open circles. 
In other words, \tzero~is enhanced by the Josephson coupling, but is suppressed by the competing order, which reduces the Josephson coupling between nearest neighbor planes\cite{chak_nat}.
If all SC orders in IPs are suppressed, i.e., $J_{\para{\rm IP}}=J_{\perp}=0$, the five-layered system is reduced to the bilayer one composed of OPs. 
Such a case is shown in Fig.~\ref{fig4} (b) by crosses, where $n$=5, $J_{\para{\rm OP}}$=1, $J_{\para{\rm IP}}$=0, $J_{\perp}$=0, and $J'_{\perp}$=0.01. 
We find that \tzero~is suppressed, but is strongly enhanced by small $K$. 
Therefore, even if the SC planes are separated by the AF insulators, the SC order can coexist with the AF order due to the Josephson coupling through the AF plane. 
The high value of \tc~in the coexistent phase is also retained by such a long-ranged Josephson coupling.

It is noted that, if one can eliminate all the \pconf~ processes, $E_J$ will be enhanced more than Eq.~(\ref{ej_afi}). 
Such a case is possible in a spin liquid state, i.e., resonating valence bond (RVB) state\cite{anderson_rvb}. 
The RVB state does not have any transition amplitude to \pconf, since the \pconf~processes corresponds to a triplet channel\cite{choi}.
Therefore, only \zconf~ process contributes to $E_J$, and then the Josephson coupling with the RVB state can be enhanced more than that with the AFI state. 

In conclusion, 
we have proposed a mechanism for high critical temperature ($T_c$) in the coexistent phase of superconducting- (SC-) and antiferromagnetic- (AF-) \cuo planes in the multilayered cuprates. The Josephson coupling between the SC planes separated by the AF plane is perturbatively calculated in terms of the hopping integral between adjacent \cuo planes. The AF interaction provides the Josephson coupling through the AF plane, which enables the coexistence and the high value of \tc~in the multilayered cuprates. 
The further enhancement of Josephson coupling is expected in a resonating valence bond state.
\begin{acknowledgments}
We would like to thank Prof. G. Baskaran, Prof. H. Fukuyama, Dr. W. Koshibae and Prof. S. E. Barnes for their valuable discussions.
 This work was supported by a Grand-in-Aid for Scientific Research on Priority Areas and the NAREGI Nanoscience Project from MEXT and CREST. One of authors (M. M.) acknowledges support by 21st Century COE program, Tohoku Univ Materials Research Center.
\end{acknowledgments}

\end{document}